\documentclass[sigconf]{acmart}
\usepackage{graphics}
\usepackage{multirow}
\usepackage{subcaption}
\usepackage{enumitem}
\usepackage{bm}
\AtBeginDocument{%
  \providecommand\BibTeX{{%
    \normalfont B\kern-0.5em{\scshape i\kern-0.25em b}\kern-0.8em\TeX}}}

\acmConference[Conference acronym 'XX]{Make sure to enter the correct
  conference title from your rights confirmation emai}{June 03--05,
  2018}{Woodstock, NY}
\begin{document}

\title{Disentangled Causal Embedding With Contrastive Learning For Recommender System}

% \author{Weiqi Zhao$^{1}$, Dian Tang$^{1}$, Xin Chen$^{1}$, Dawei Lv$^{1}$, Daoli Ou$^{1}$, Biao Li$^{1}$, Peng Jiang$^{1\ast}$, Kun Gai$^{2}$
% }
% % \authornote{Corresponding Author}
% \affiliation{%
%   \institution{$^1$Kuaishou Technology \city{Beijing} \country{China};  $^2$Unaffiliated} 
%   \institution{$\ast$Corresponding Author: jp2006@139.com}
%   % \city{Beijing}
%   % \country{China}
% }
% \email{{zhaoweiqi03,tangdian,chenxin,lvdawei,oudaoli,libiao,jiangpeng}@kuaishou.com}
% % \affiliation{%
% %  \institution{Institute for Clarity in Documentation}
% %  \streetaddress{P.O. Box 1212}
% %  \city{Dublin}
% %   \state{Ohio}
% %  \country{USA}
% %  \postcode{43017-6221}
% %}

\author{Weiqi Zhao}
\affiliation{%
  \institution{Kuaishou Technology} 
  \city{Beijing}
  \country{China}
}
\email{zhaoweiqi03@kuaishou.com}

\author{Dian Tang}
\affiliation{%
  \institution{Kuaishou Technology} 
  \city{Beijing}
  \country{China}
}
\email{tangdian@kuaishou.com}

\author{Xin Chen}
\affiliation{%
  \institution{Kuaishou Technology} 
  \city{Beijing}
  \country{China}
}
\email{chenxin@kuaishou.com}

\author{Dawei Lv}
\affiliation{%
  \institution{Kuaishou Technology} 
  \city{Beijing}
  \country{China}
}
\email{lvdawei@kuaishou.com}

\author{Daoli Ou}
\affiliation{%
  \institution{Kuaishou Technology} 
  \city{Beijing}
  \country{China}
}
\email{oudaoli@kuaishou.com }

\author{Biao Li}
\affiliation{%
  \institution{Kuaishou Technology} 
  \city{Beijing}
  \country{China}
}
\email{libiao@kuaishou.com}

\author{Peng Jiang}
\authornote{Corresponding author.}
\affiliation{%
  \institution{Kuaishou Technology} 
  \city{Beijing}
  \country{China}
}
\email{jiangpeng@kuaishou.com}

\author{Kun Gai}
\affiliation{%
  \institution{Unaffiliated} 
  \city{Beijing}
  \country{China}
}
\email{gai.kun@qq.com}

\renewcommand{\shortauthors}{Weiqi Zhao, et al.}

\begin{abstract}
Recommender systems usually rely on observed user interaction data to build personalized recommendation models, assuming that the observed data reflect user interest. However, user interacting with an item may also due to conformity, the need to follow popular items. Most previous studies neglect user's conformity and entangle interest with it, which may cause the recommender systems fail to provide satisfying results. 
% Other studies extract pure interest by eliminating popularity bias, like unified re-weighting instances to decrease the impact of popular items, which ignores user’s conformity and hurts user experience. 
Therefore, from the \textit{cause-effect} view, disentangling these interaction \textit{causes} is a crucial issue. It also contributes to OOD problems, where training and test data are out-of-distribution. Nevertheless, it is quite challenging as we lack the signal to differentiate interest and conformity. The data sparsity of pure cause and the items' long-tail problem hinder disentangled causal embedding. In this paper, we propose DCCL, a framework that adopts contrastive learning to disentangle these two causes by sample augmentation for interest and conformity respectively. Futhermore, DCCL is model-agnostic, which can be easily deployed in any industrial online system. Extensive experiments are conducted over two real-world datasets and DCCL outperforms state-of-the-art baselines on top of various backbone models in various OOD environments. We also demonstrate the performance improvements by online A/B testing on Kuaishou, a billion-user scale short-video recommender system.
\end{abstract}






\maketitle

\section{Introduction}
\begin{figure}[t]
  \centering
  \setlength{\abovecaptionskip}{-0.001cm}
  \setlength{\belowcaptionskip}{-0.55cm}
  \includegraphics[width=0.98\linewidth]{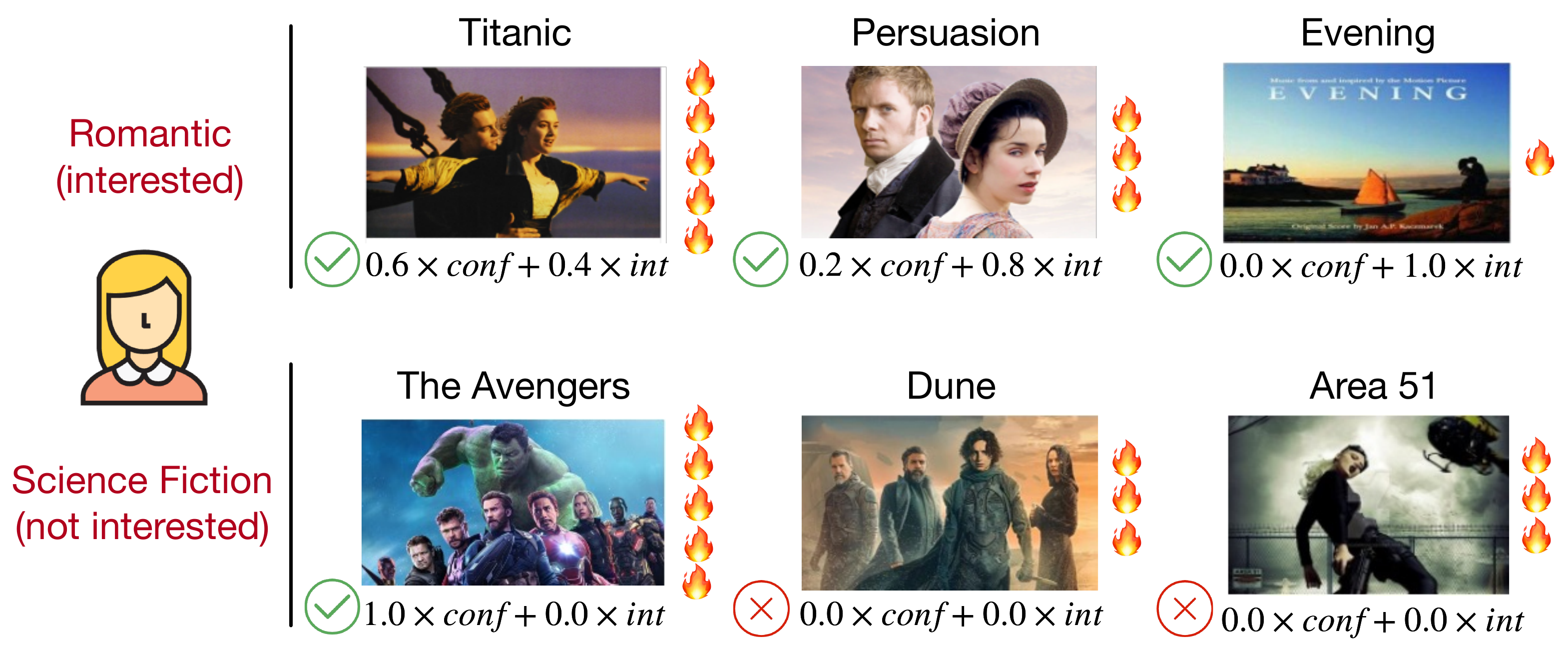}
  \caption{Examples of interactive intention for different items. Int and Conf indicate the interest and conformity, respectively.}
  \label{fig:intention}
\end{figure}

In recent years, recommender systems (RS) have been widely used in countless online applications, such as e-commerce \cite{zhou2018deep, ying2018graph, wu2019neural}, social media \cite{lei2020estimation,he2017neural} and digital streaming \cite{kabbur2013fism, wei2019mmgcn}, etc., which provide users with personalized contents by mining user preference from the user-item interaction data \cite{perc2014matthew, pi2020search, Cheating2021clicks, yang2021top, Nondisplay2019improving, denoising2021}.
% Specifically, A large number of models trained directly on the observational interaction dataset are deployed online to maximize the likelihood that a user clicks the recommended items \cite{zhou2018deep, kabbur2013fism, pi2020search}. 
User interest is absolutely an important reason for item interaction. But users may also click items due to conformity, the need to follow popular items. Therefore, the observed interaction data is attributed to two \textit{causes}, interest and conformity simultaneously. As an example shown in Figure \ref{fig:intention}, a girl who's mainly interested in Romantic movies also intends to watch The Avengers, one of the most popular movies. Although she's not interested in Science Fiction, it attracts her to figure out why The Avengers can receive so much attention. Besides, her every choice consists of interest and conformity and the distribution varies from different movies. 
% Even in Romantic movies, watching Evening is mainly for interest, while choosing Titanic has 60\% conformity intention. 
This example demonstrates that conformity is also an inherent need that deserves full attention as well as interest. 

However, most previous studies ignore the value of conformity. Since it interferes the interest modeling, conformity has been long considered harmful. A lot of studies eliminate popularity bias to extract pure interest \cite{abdollahpouri2017controlling, PDA2021causal, mansoury2020feedback}, e.g. based on inverse propensity scoring (IPS) \cite{gruson2019offline, schnabel2016recommendations, wang2018deconfounded}, or leverage a small fraction of unbiased data \cite{bonner2018causal}.
% \cite{MACR21} analyzes the influence of popularity from the perspective of causal mechanism and performs counterfactual inference to eliminate popularity bias. 
These operations remove conformity effect and only recommend items based on interest. \cite{PDA2021causal} realizes the popularity has positive effect which should be maintained. But the popularity effect of an item is the same for all users, which does not reflect the conformity's personalization. Furthermore, popularity is usually \textbf{O}ut-\textbf{O}f-\textbf{D}istribution (OOD) in training and test data in industrial applications, and the above methods cannot effectively solve it. \cite{DICE21} learns disentangled interest and conformity embeddings on both user and item sides by splitting observed data into conformity only and integrated parts according to the relative popularity of item pairs. It illustrates the necessity of disentangling the \textit{causes} (interest and conformity) for the \textit{effect} (interaction): models are more robust and interpretable with disentangled causal embedding. However, challenges of disentangling causal embedding still remain: First, interest and conformity are integrated in observed data and we lack the signal to distinguish them, making it difficult to train on raw observed data. Second, the interactions with different causes are not sufficient, and a large number of long-tail items exist in real-world recommender systems. These two sparsity problems make disentangled causal embeddings harder to learn.

% \begin{itemize}[leftmargin=*]
% \item First, interest and conformity are integrated in observed data and we lack the signal to distinguish them. The lack of ground-truth makes it challenging to train disentangled causal embedding based on raw observed interaction data.
% \item Second, there are two data sparsity problems with the observed data. The interactions with different causes are not sufficient. Besides, a large number of long-tail items exist in real-world recommender systems. These two sparsity problems make disentangled causal embeddings harder to learn.
% \end{itemize}

\begin{figure}[t]
    \centering
    \setlength{\belowcaptionskip}{-0.6cm} 
    \includegraphics[width=0.7\linewidth]{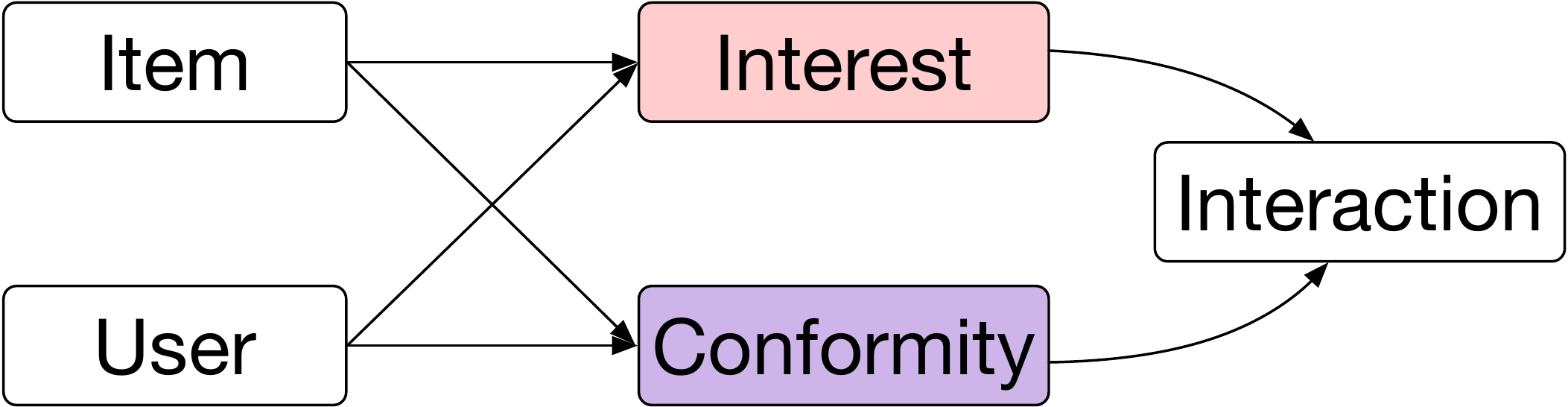}
    \caption{Causal graph and embeddings. We make concise causal modeling on each interaction which is the combination of interest and conformity.}
    \label{fig:causal graph}
\end{figure}

In this work, we first formulate a causal graph \cite{pearl2009causality, bookcausal2017} (Figure \ref{fig:causal graph}) to describe the interaction causes. Thus, the interaction of each user-item pair is composed of two causes: interest and conformity. Both causes are affected by user and item simultaneously. Then we propose a general framework for \textbf{D}isentangling \textbf{C}ausal Embedding with \textbf{C}ontrastive \textbf{L}earning (DCCL) to learn disentangled interest and conformity embeddings based on causal mechanism \cite{book-causal2020, bookcausal2017}. The framework only deals with embeddings and can be applied on any specific models, such as MF and LightGCN. Contrastive learning \cite{zhou2021contrastive, Sequencecontrastive2020contrastive} is a self-supervised technique used to enhance model representation by augmenting samples in similar and different views, which is very effective against data sparsity. To better learn disentangled embeddings, we augment samples for interest and conformity respectively by leveraging the popularity signal. 
To summarize, the main contributions of this paper are as follows:
\begin{itemize}[leftmargin=*]
\item We propose a general framework to learn disentangled causal embeddings with contrastive learning based on observed data directly. The sample augmentation provides sufficient sample for different causes and long-tail items, which can effectively deal with data sparsity problems and enhance cause representations. Futhermore, the framework can be applied to any base model.

\item Extensive experiments are conducted on two real-world datasets. Results show that our method achieves significant improvements over SOTA baselines, and more robust for OOD environments.
% Further analysis demonstrates that our method successfully captures user preference and the distribution over two causes, and the learned causal embeddings possess high interpretability. 
We deploy DCCL on Kuaishou, a billion-user scale short-video recommender system. The online A/B testing shows great improvements in effective-view-through rate and like-through rate, especially for long-tail items.
\end{itemize}

\section{METHODOLOGY}

\subsection{Causal Embedding for Recommendation}
With the goal of strengthening recommender systems, several causal related methods have been proposed, such as IPS \cite{gruson2019offline, schnabel2016recommendations}, causal embedding \cite{DICE21, bonner2018causal} and counterfactual inference \cite{MACR21, Cheating2021clicks}, etc. In this paper, we propose a framework based on causal embedding which formulates a causal graph to describe the important cause-effect relations in the recommendation process in Figure \ref{fig:causal graph}. Disentangling each cause into causal embedding has the follow advantages. On the one hand, it models the user's personalized preference accurately for different causes from a view of interaction generation. On the other hand, causal modeling can lead to more robust models, with stronger generalization ability.

Moreover, there are usually multiple causes for the user-item interaction, such as item popularity, category, and quality, etc. Following previous work \cite{DICE21}, we focus on these two causes: interest and conformity to model the user-item interaction.

\begin{figure}[t] 
    \centering
    \vspace{-0.8cm}
    \setlength{\belowcaptionskip}{-0.5cm}
    \includegraphics[width=1.0\linewidth,scale=0.01]{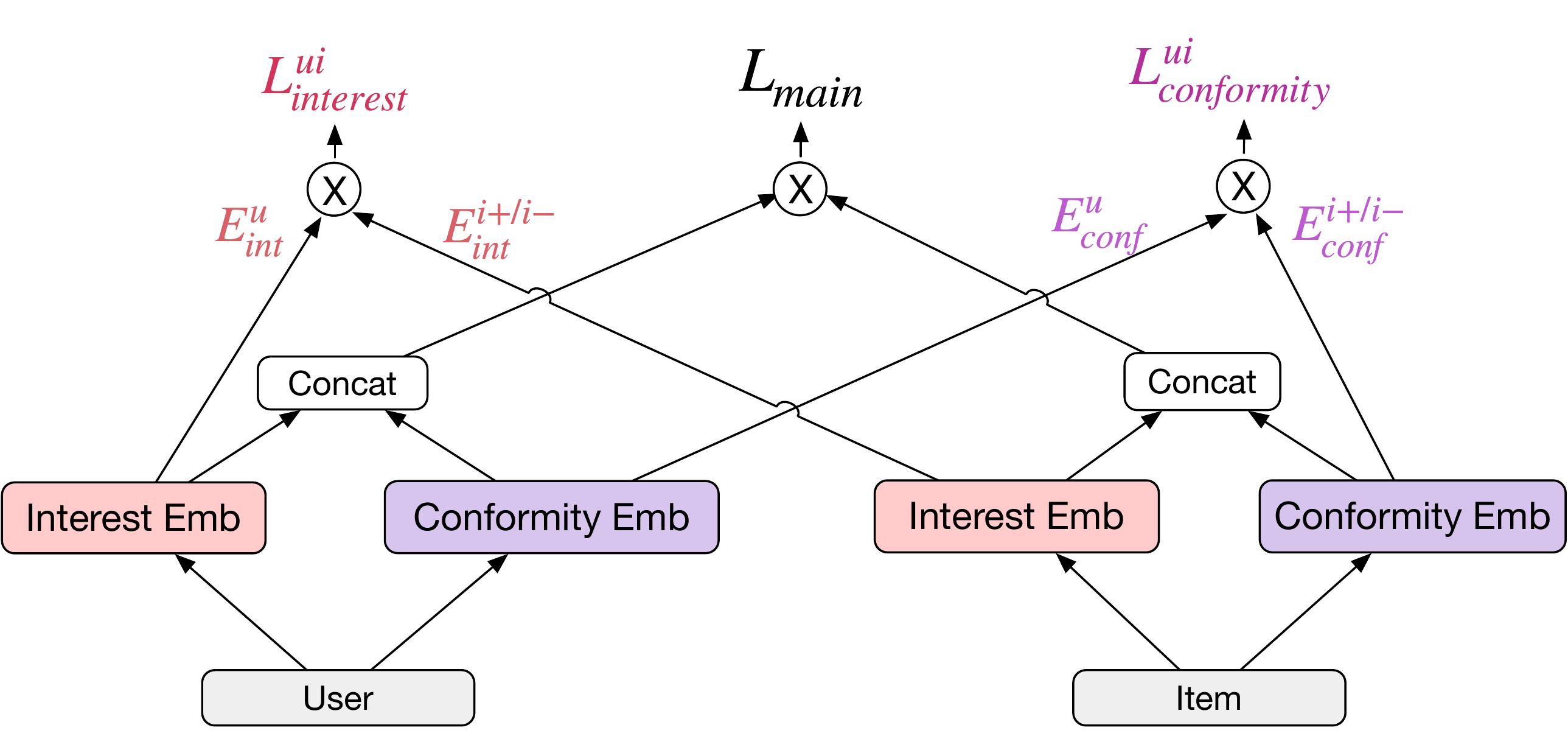}
    \caption{The framework of DCCL. $E_{int}^{i+/i-}$ and $E_{conf}^{i+/i-}$ represent the item embeddings of augmented interest and conformity samples, respectively}
    \label{fig:overall}
\end{figure}

\subsection{Contrastive Learning}
In real-world recommender systems, there are a large number of long-tail items, and the interactions are extremely sparse. These two sparsity problems make disentangled embeddings harder to learn. Therefore, to ensure sufficient learning of disentangled cause embeddings on observed interaction data directly, we utilize the contrastive learning to augment samples for each cause.

% \subsubsection{User-Item Pair Contrastive Learning}\label{UI pair for Contrastive Learning}
The overall structure of proposed framework is shown as Figure \ref{fig:overall}.
Two user-item pair contrastive learning tasks are designed to learn the interest and conformity embeddings, respectively. We define $E_{int}^u$ and $E_{conf}^u$ as the disentangled interest and conformity embeddings for user $u$. For an item $i$, the causal embedding $E_{int}^i$ is related to the item content, $E_{conf}^i$ is related to the item popularity. Specifically, for a given mini-batch with $N$ samples, we regard the interacted items of user $u$ are all positive samples and the items from other users are regarded as negative samples for user $u$.

We represent the disentangled interest causal embedding of positive pair as $(E_{int}^{u}, E_{int}^{i+})$, and the negative pair as $(E_{int}^{u}, E_{int}^{i-})$. Besides, the dot product function $S(w,v) = w^\top v$ is adopted to measure the similarity between each pair of representations. In addition, it is necessary to integrate the popularity signal of the target item into contrastive loss to ensure that interaction with long-tail items is principally based on full interest. Therefore, the user-item pair contrastive loss for interest is defined as follows:

\begin{equation}
\setlength{\abovedisplayskip}{-2pt}
\begin{aligned}
&\mathcal{L}_{int}^{ui} = \\
&    -\frac{1}{N}\sum log\frac{exp(-I_{pop}) \times exp(S(E_{int}^{u} ,E_{int}^{i+}))}{exp(S(E_{int}^{u},E_{int}^{i+}))
    + \sum exp(S(E_{int}^{u}, E_{int}^{i-}))},
\label{equ ui interest}
\end{aligned}
\end{equation}
where $I_{pop}$ represents the normalized popularity of the interacted item, which is defined as the ratio of the number of the item user interacted vs. total item impression number. For the high popular items, the popularity weight $exp (-I_{pop})$ is close to $exp(-1)$ rather than 0, which enables the task to learn interest preference to a certain extent.
Integrating the popularity signal into contrastive loss of interest embedding pair helps to directly learn disentangled interest representations from interaction data.

Similarly, for conformity causal embedding, $(E_{conf}^u$, $E_{conf}^{i+})$ is the positive user-item pair, and $(E_{conf}^u, E_{conf}^{i-})$ is the negative pair. Note that for augmented negative samples of conformity, we filter out the ones with higher popularity than target item to ensure that the current interaction is primarily based on the user's conformity to popular target item. The popularity signal of target item is also integrated into user-item pair contrastive loss for conformity as follows:

\begin{equation}
\setlength{\abovedisplayskip}{-2pt}
\begin{aligned}
&\mathcal{L}_{conf}^{ui} = \\
&    -\frac{1}{N}\sum log\frac{ (1 - exp(-I_{pop})) \times exp(S(E_{conf}^u,E_{conf}^{i+}))}{exp(S(E_{conf}^u,E_{conf}^{i+}))
    + \sum exp(S(E_{conf}^u, E_{conf}^{i-}))},
\label{equ conformity}
\end{aligned}
\end{equation}
where $1 - exp(-I_{pop})$ ensures interaction with higher popularity item is attributed more to conformity.
This contrastive loss updates model parameters by minimizing the deviation of positive examples and maximizing the relevance of augmented negative samples. The careful negative sample augmentation and popularity weight enables us to learn disentangled conformity embedding well.

\subsection{Learning and Discussion}
We leverage the multi-task training strategy to optimize the main and two contrastive learning tasks jointly as follows:

\begin{equation}
\setlength{\abovedisplayskip}{-2pt}
    \mathcal{L}_{total} = \mathcal{L}_{main} + \alpha \mathcal{L}_{int}^{ui} + \beta \mathcal{L}_{conf}^{ui},
\end{equation}
where $\mathcal{L}_{main}$ represents the main loss of recommendation. $\alpha$ and $\beta$ are the hyper-parameters to balance these three loss. 

How to understand the effect of each auxiliary task and the relationship between them? Intuitively, we assume the interaction with a popular item is mainly attributed to conformity, while interaction with a long-tail item is mainly due to interest. Therefore, as popularity increases, the popularity weight of $\mathcal{L}_{int}^{ui}$ in Equ. \ref{equ ui interest} decreases, while the weight of $\mathcal{L}_{conf}^{ui}$ in Equ. \ref{equ conformity} increments. This assumption effectively directs model disentangling interest and conformity, but in case users interact with popular items due to great interest, the interest embedding cannot be well represented. 

Similar to DICE \cite{DICE21}, our method also applies popularity as a supervised signal to learn the disentangled causal embeddings. However, our method directly learns causal embeddings on the raw interaction data instead of dividing different training sets to learn corresponding causal embedding according to item popularity. This hard division method may introduce noise and aggravate the problem of data sparsity.

\section{EXPERIMENTS}
In this section, we conduct experiments to evaluate the performance of the proposed framework.
\subsection{Experiment Settings}
{\bfseries Datasets.} We conduct experiments on two real-world datasets:
% The first two datasets contains tens of millions of interactions between millions of items and users. 
Yelp $\footnote{\url{https://www.yelp.com/dataset}}$, and an industrial Short-video dataset   $\footnote{\url{https://github.com/somestudies/DCCL}}$ which is collected from a large-scale short video stream platform. 
% Following the common practice in \cite{Sequencecontrastive2020contrastive, zhou2020s3},
we convert all numeric ratings, click behavior to “1” and others to “0”, and take the 10-core version for experiments as \cite{cheng20183ncf, zhou2020s3}, where users and items with fewer than 10 interaction records are discarded. The statistics of these two datasets after pre-processing are summarized in Table \ref{tab:datesets}.

{\bfseries Experiment Setups.}
Following \cite{DICE21}, causal approaches usually serve as additional methods upon backbone recommendation models. Thus, we select the most commonly adopted recommendation model MF \cite{rendle2012bpr}, and graph-based model LightGCN \cite{he2020lightgcn} as backbone models to compare different approaches. In terms of the parameters setting, the embedding size $d$ is fixed as 64 and the batch size $B$ is 512 and the model learning rate $\gamma$ is 0.001. 
% Based on the fact that the magnitude of historical sequence contrastive loss is larger than user-item pair contrastive loss, 
We set $\alpha = 0.1$ and $\beta = 0.1$ which achieves great performance in experiments. 
For a fair comparison, we apply Bayesian Personalized Ranking \cite{rendle2012bpr} (BPR) loss  as the main loss for all methods. 
Since our DCCL framework introduces extra costs for calculating contrastive loss in training procedure, the complexity of our framework in each batch is $\mathcal{O}(B^2d)$ compared to the $\mathcal{O}(Bd)$ of the original backbone model. 
Meanwhile, the model inference procedure of DCCL is the same with other framework. 
Therefore, the time complexity does not change during inference.
\begin{table}
    \centering
    \caption{Statistics of datasets after pre-processing.}
    \begin{tabular}{c|c|c|c}
    \hline
    Dataset & \#Users & \#Items  & Sparsity \\
    \hline
      Yelp & 27057 & 17843 & $1.007 \times 10^{-3}$ \\ 
      Short-video & 30957 & 71006 & $1.145 \times 10^{-3}$ \\
     \hline
    \end{tabular}
    \label{tab:datesets}
\end{table}

\subsection{Performance Comparison}
For offline evaluation, we compare DCCL with several SOTA causal recommendation methods including {\bfseries IPS} \cite{schnabel2016recommendations}, {\bfseries CausE} \cite{bonner2018causal}, {\bfseries DICE} \cite{DICE21}, {\bfseries PD} \cite{PDA2021causal}, {\bfseries MACR} \cite{MACR21}.
We evaluate top-$K$ recommendation performance based on the metrics including Hit Ratio (HR) and Normalized Discounted Cumulative Gain (NDCG). As shown in Table \ref{tab:results}, proposed DCCL consistently outperforms all baselines with significant improvements with respect to all metrics on two datasets, which demonstrates the effectiveness of DCCL. The average improvement of DCCL$\_$MF over MF is 33.24\%,  and DCCL$\_$LightGCN over LightGCN is 22.91\% in terms of HR@20 on both datasets.

Compared with primitive MF and LightGCN, CausE \cite{bonner2018causal} obtains improvement by eliminating the bias in interaction data. However, the improvement effect is limited because conformity preference is not considered. The effect of DICE \cite{DICE21} has not been greatly improved because the data sparsity makes it difficult to achieve sufficient learning of causal embedding. 
Furthermore, PD \cite{PDA2021causal} and MACR \cite{MACR21} achieve better results compared to other baselines by leveraging the conformity (popularity) in their models. However, thanks to contrastive learning for more accurate causal embeddings, our method outperforms them over 10\% in most cases.

To investigate the integral effects of two auxiliary tasks including interest pair contrastive learning (IPCL) and conformity pair contrastive learning (CPCL), we conduct ablation studies on the two datasets and remove different components at a time for comparisons. As shown in Table \ref{tab:results}, the results validate the significant improvement brought by designed contrastive learning tasks. Compared with removing the other two auxiliary tasks, the user-item interest auxiliary task has more influence on the final effect due to the purer embedding of interest after disentangling. Specially, these two special cases also outperform all the baselines with HR@20 and NDCG@20 on the two datasets, which further illustrates the effectiveness of contrastive learning auxiliary tasks.

\begin{table}[t]
\centering
    \setlength{\belowcaptionskip}{-0.22cm} 
    \caption{Overall performance on two real-world datasets with $K$ = 20. The best results are highlighted in bold, and IPCL, CPCL refer to the user-item interest pair and user-item conformity pair contrastive learning tasks, respectively.}
    \begin{tabular}{c|c|c c |c c}
    \hline
    ~ & ~ &  \multicolumn{2}{|c|}{Yelp} & \multicolumn{2}{|c}{Short-video} \\
    \hline
    Backbone &Method & HR & NDCG & HR & NDCG \\
    \hline
    \multirow{9}*{MF} & - & 5.070 & 1.951  & 2.613 & 1.018 \\ 
    ~ & CausE & 5.072 & 1.973 & 2.629 & 1.021 \\ 
    ~ & IPS & 4.822 & 1.917  & 2.235 & 0.879 \\
    ~ & DICE & 5.172 & 2.044 & 2.645 & 1.034 \\ 
    ~ & PD & 5.183 & 2.062 & 2.754 & 1.076 \\
    ~ & MACR & 5.224 & 2.104 & 2.867 & 1.137\\
    ~ & DCCL w/o IPCL & 5.431 & 2.143 & 3.202 & 1.241\\
    ~ & DCCL w/o CPCL & 5.607 & 2.216 & 3.489 & 1.343\\
    ~ & DCCL & \textbf{6.171} & \textbf{2.415}  & \textbf{3.783} & \textbf{1.463}\\
    \hline
    \multirow{9}*{LightGCN} & - & 5.944 & 2.372  & 3.727 & 1.476 \\
    ~ & CausE & 5.958 & 2.352 & 3.827 & 1.502\\
    ~ & IPS & 5.781 & 2.304 & 3.606 & 1.421 \\
    ~ & DICE & 6.375 & 2.498 & 4.064 & 1.603 \\
    ~ & PD & 6.451 & 2.514 & 4.106 & 1.609 \\
    ~ & MACR & 6.573 & 2.552 & 4.142 & 1.613 \\
    ~ & DCCL w/o IPCL & 6.717 & 2.672 & 4.151 & 1.626 \\
    ~ & DCCL w/o CPCL & 6.821 & 2.705 & 4.311 & 1.653 \\
    ~ & DCCL & \textbf{7.254} & \textbf{2.816} & \textbf{4.613} & \textbf{1.713}\\
    \hline
    \end{tabular}
    \label{tab:results}
\end{table}

% \end{itemize}
\subsection{Evaluation of OOD Prediction}
In order to evaluate the performance of causal learning under OOD environments, intervened test sets are needed. We define item whose popularity is greater than the 80th percentile as popular item. Based on this definition, the proportion of popular items of the original Short-video test set accounts for 60\%, which is consistent with the distribution of the training set. We sample different proportions of popular items from Short-video dataset and construct three extra intervened test sets, where the proportions of popular items are 50\%, 40\% and 30\%, respectively. 

\begin{figure}
  \centering
  \vspace{-0.3cm} 
  \setlength{\abovecaptionskip}{0.05cm}       
  \setlength{\belowcaptionskip}{-0.3cm} 
  \includegraphics[scale=0.26]{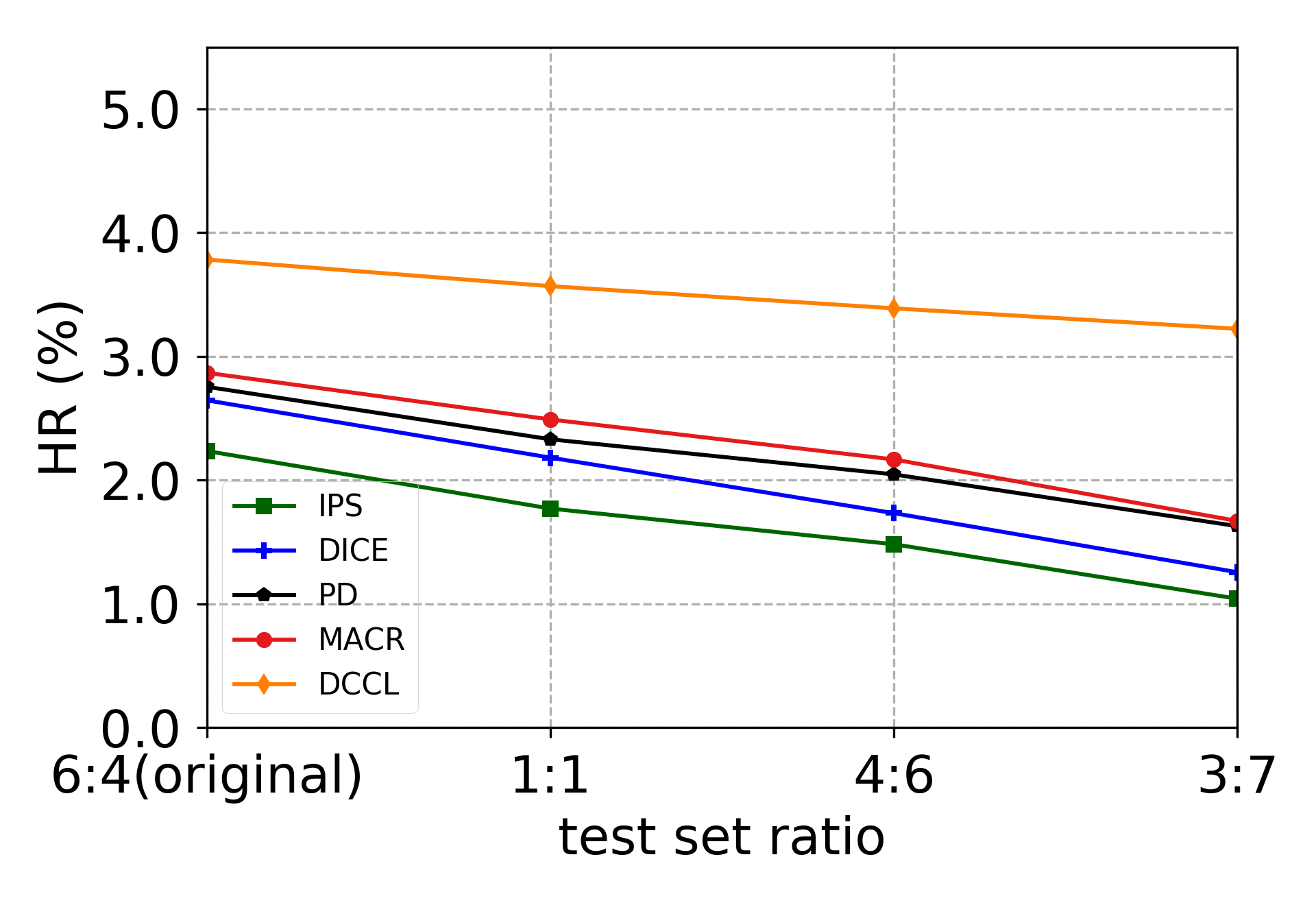}
  \hspace{-0.1in}
  \includegraphics[scale=0.26]{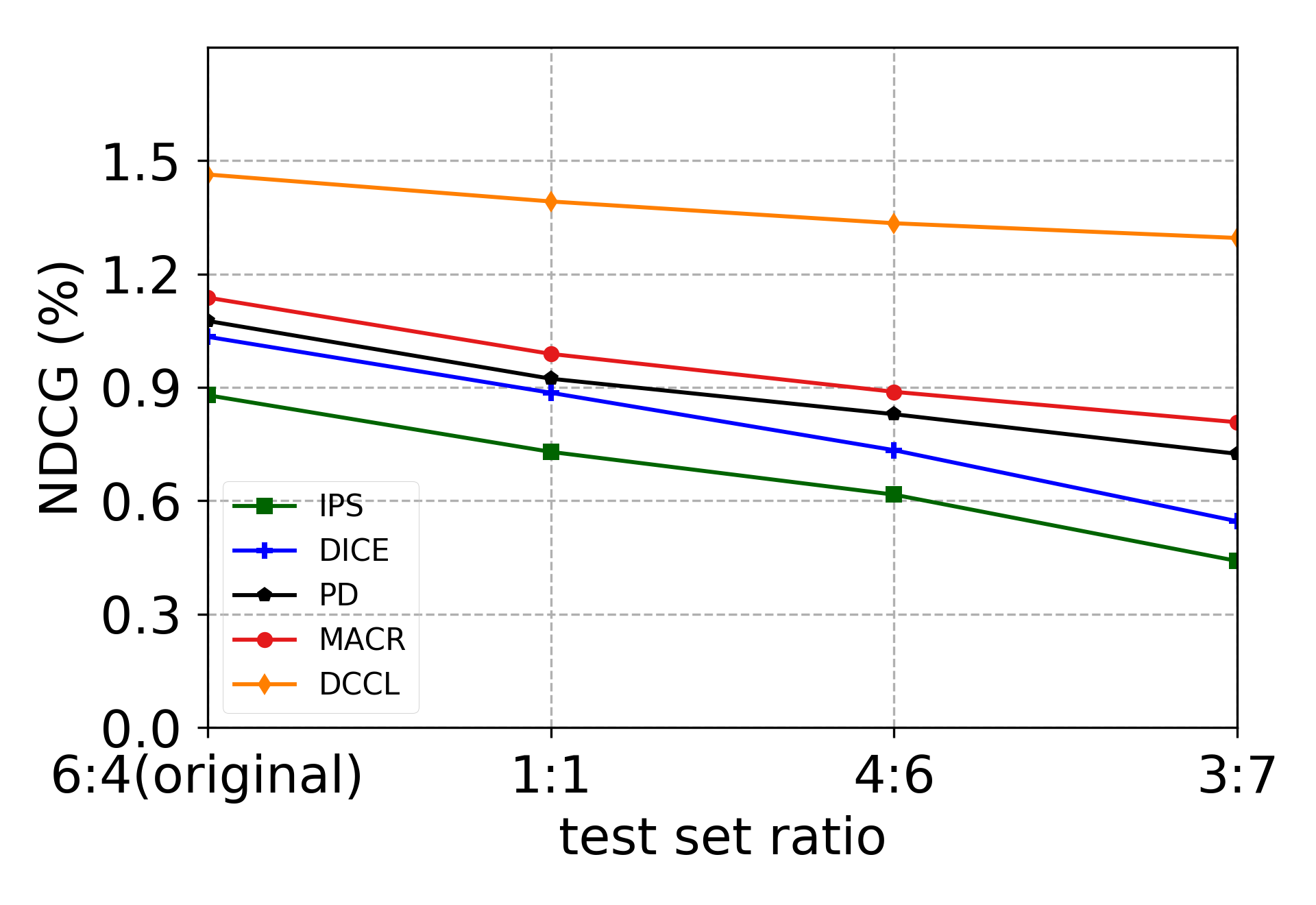}
  \caption{Effect of DCCL and other baselines on Short-video dataset with distribution shift.}
   \label{fig:distribution_shift_prediction}
\end{figure}

We evaluate the HR@20 and NDCG@20 of DCCL and other baselines on these intervened test sets, and the results are shown in Figure \ref{fig:distribution_shift_prediction}. We observe that the metrics for all methods decrease when the distribution of the sampled test set differs from that of the original test set. However, DCCL learns causal embeddings more accurately on both user and item sides. Our method outperforms MACR by over 32\% on the original test set and over 93\% on the sampled test set, in which the proportion of popular items is 30\%. As the popularity distribution shift increases, the advantage of DCCL becomes more significant, verifying that DCCL can certainly promote stability and stronger robustness in various OOD environments.

\subsection{Online Experiments}
In this subsection, online A/B experiments are conducted on Kuaishou App $\footnote{\url{https://www.kuaishou.com/new-reco}}$, a billion-user scale short-video recommender system, to verify the superior performance of proposed framework DCCL in industrial recommendation scenarios. 

The practical implementation details for online experiments are as follows. We use user id as user-side features, and video id as item-side feature. The samples with user interactions, i.e. effective-view (watching video time exceeds a threshold) and like, are regarded as positive ones. Negative samples are randomly selected from the videos that the user doesn't interact with. We use MF as the backbone model and implement DICE, MACR and DCCL upon it. DICE and MACR are selected because they're the top baselines in offline experiments. All models are deployed for retrieval modules. As for training parameters, and the dimension of embedding for both user and item is 64. Besides, we use Adam optimizer with mini-batch size 8196 and learning rate 0.001.

\begin{figure}
  \centering
  \vspace{-0.2cm} 
  \setlength{\abovecaptionskip}{-0.003cm} 
  \setlength{\belowcaptionskip}{-0.6cm} 
  \includegraphics[scale=0.25]{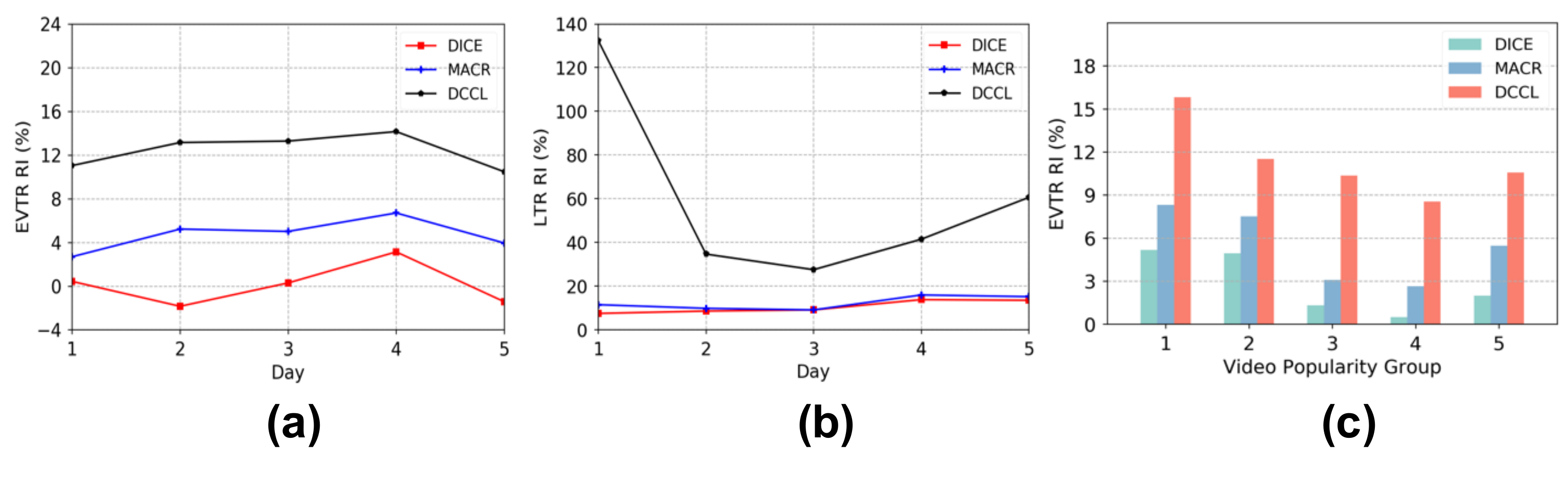}
  \hspace{-0.18in}
  \caption{Online A/B results of EVTR and LTR. “RI” refers to the relative improvement vs MF.}
   \label{fig:online results}
\end{figure}

The online experiment lasts for five days. We focus on the metrics of Effective-View-Through Rate (EVTR) and Like-Through Rate (LTR), which reflect user satisfaction with recommended videos. Specifically, EVTR is a statistic of an average rate of the number of videos that watch time exceeds a threshold vs. video impression number. In addition, LTR is defined as average rate of the number of videos user likes vs. video impression number.

As depicted in Figure \ref{fig:online results} (a) and Figure \ref{fig:online results} (b), our method performs the best recommendation compared with other baselines. Compared to MACR, the overall average relative improvements are 7.362\% with EVTR and 41.82\% with LTR respectively, which is very significant in our online system. DCCL outperforms DICE due to the fact that we apply contrastive learning for sample augmentation to update disentangled embeddings, which better alleviates the sparsity problem. For long-tail performance, Figure \ref{fig:online results} (c) shows the EVTR improvement under different popularity group of videos. The great improvement of unpopular video group verifies that our approach is more friendly to long-tail videos. Overall, the online experiment demonstrates the effectiveness of DCCL framework in large-scale industrial recommendation scenarios.

\section{CONCLUSION}
In this paper, we propose a general framework DCCL to learn disentangled causal embeddings based on causal mechanism and causal graph, representing a fine-grained analysis to better understand how interest and conformity affect the recommendation process. 
% This work shows the advantages of disentangled causal embeddings in combination with contrastive learning, which makes it possible for the causal embedding methods to be directly trained on the observed interaction data without addition unbiased data or policy data. We believe that how to construct finer level constractive task for underlying causes will further improve the recommendation effect in the future.
% This work shows the advantages of disentangled causal embeddings in combination with contrastive learning, which makes it possible for the causal embedding methods to be directly trained on the observed data. 
We believe that how to construct finer-level contrastive task for underlying causes will further improve the recommendation effect in the future.

% \section{Appendices}

% If your work needs an appendix, add it before the
% ``\verb|\end{document}|'' command at the conclusion of your source
% document.

% Start the appendix with the ``\verb|appendix|'' command:
% \begin{verbatim}
%   \appendix
% \end{verbatim}
% and note that in the appendix, sections are lettered, not
% numbered. This document has two appendices, demonstrating the section
% and subsection identification method.

% \section{SIGCHI Extended Abstracts}

% The ``\verb|sigchi-a|'' template style (available only in \LaTeX\ and
% not in Word) produces a landscape-orientation formatted article, with
% a wide left margin. Three environments are available for use with the
% ``\verb|sigchi-a|'' template style, and produce formatted output in
% the margin:
% \begin{itemize}
% \item {\verb|sidebar|}:  Place formatted text in the margin.
% \item {\verb|marginfigure|}: Place a figure in the margin.
% \item {\verb|margintable|}: Place a table in the margin.
% \end{itemize}

% \begin{acks}
% \end{acks}

\bibliographystyle{ACM-Reference-Format}
\bibliography{main}

% \appendix









\end{document}